\documentclass[aps,preprint,prd,showpacs,nofootinbib]{revtex4}
\usepackage{amsmath}
\usepackage{graphicx}
\usepackage{dcolumn}
\usepackage{bm}
\usepackage{amssymb}
\usepackage{latexsym}
\usepackage{color}

\def\be{\begin{equation}}
\def\ee{\end{equation}}
\def\ba{\begin{eqnarray}}
\def\ea{\end{eqnarray}}

\begin{document}

\title{Obtaining the CMB anomalies with a bounce from the contracting phase to inflation  }

\author{Zhi-Guo Liu$^{1,}$\footnote{Email: liuzhiguo08@mails.ucas.ac.cn}}
\author{Zong-Kuan Guo$^{2,}$\footnote{Email: guozk@itp.ac.cn}}
\author{Yun-Song Piao$^{1,}$\footnote{Email: yspiao@ucas.ac.cn}}

\affiliation{$^1$ School of Physics, University of Chinese Academy of
Sciences, Beijing 100049, China}

\affiliation{$^2$ State Key Laboratory of Theoretical Physics, Institute of Theoretical Physics, \\
Chinese Academy of Sciences, P.O. Box 2735, Beijing 100190, China}

\begin{abstract}

Recent Planck data show the anomalies of CMB fluctuations on large
angular scales, which confirms the early observations by WMAP. We
continue studying an inflationary model, in which before the slow
roll inflation the universe is in a contracting phase, and fit the
model with the Planck data. It is showed that this model may
generate not only the power deficit at low-$l$, but also a large
hemispherical power asymmetry in CMB. We also discuss the
implication of the result to the eternal inflation scenario.

\end{abstract}

\maketitle

\section{Introduction}

The inflation scenario is the current paradigm of the early
universe. The inflation may be realized with the inflationary
models, which will be identified by the observations. Recently,
the Planck collaboration has released the data of the power on
cosmic microwave background (CMB)
\cite{Ade:2013xsa},\cite{Ade:2013uln}, which prefers the single
field slow roll inflationary model with a concave potential.

However, the Planck collaboration has reported a power deficit in
the low-$l$ CMB power spectrum at $l\lesssim 40$
\cite{Ade:2013nlj}, which also is found in WMAP data, and not
concordant with the Planck bestfit model, though the data points
are still consistent well with the cosmic variance. Its
statistical significance is about $2.5\sim 3\sigma$. In the
meantime, the Planck collaboration has also reported a
hemispherical power asymmetry in CMB \cite{Ade:2013nlj}, which
conformed a similar result of WMAP \cite{Eriksen:2007pc,Hoftuft:2009rq}, but has
better precision. The Planck data have larger statistical
significance than in the WMAP data, which makes this asymmetry
difficult to attribute the asymmetry to foregrounds.

These anomalies are actually intriguing, which has motivated some
relevant studies. The curvaton scenario may explain the power
asymmetry
\cite{Erickcek:2008sm},\cite{Lyth:2013vha},\cite{Wang:2013lda},\cite{McDonald:2013aca},\cite{Namjoo:2013fka}.
However, we will consider a different possibility, i.e. these
anomalies might be a hint of the preinflationary physics. In this
case, the inflation might last for just the minimum number of
efoldings, the Planck bestfit single field inflationary model only
actually provides a fit for the intermediate and small angular
scales. After the WMAP1 data, the power deficit at low-$l$ has
been investigated in
Refs.\cite{Contaldi:2003zv},\cite{Piao:2003zm},\cite{Piao:2003hh},\cite{Powell:2006yg},\cite{Boyanovsky:2006qi},\cite{Mielczarek:2008pf},\cite{Mortonson:2009xk}
\cite{Liu:2010fm},\cite{Dudas:2012vv},\cite{BouhmadiLopez:2012by}
along this line.

The study of the bouncing model has a long history, e.g.,Pre
big-bang (PBB) scenario \cite{MGV} and ekpyrotic scenario
\cite{Khoury:2001wf}. In the bouncing model, initially the
universe is in a contracting phase, and then it bounces into an
expanding phase, which results in a solution to the cosmological
singularity problem. In
Refs.\cite{Piao:2003zm},\cite{Piao:2003hh}, the model, in which
before the slow roll inflation the universe is in a contracting
phase and after the bounce it begins to inflate, has been studied,
which will be called the bouncing inflation model for simplicity
here. In this model, the contracting phase is similar to that in
PBB scenario, see \cite{GV2},\cite{LWC} for reviews, and in
principle also may be that in ekpyrotic scenario. In the PBB
scenario, the spectrum of the adiabatic perturbation generated
during the kinetic contraction is highly blue, which is not
consistent with the observations. However, here this blue spectrum
is just required for the power suppression on large angular scale
\cite{Piao:2003zm}.

The slow roll inflation generally start in a high scale, which is
required to insure that the amplitude of primordial perturbation
is consistent with the observations and the reheating temperature
is suitable with a hot big bang evolution after inflation.
Recently, in the eternal inflation scenario, it has been argued
that if the scale of the eternally inflating background is very
low, the beginning of the slow roll inflation will requires a
large uptunneling, which is exponentially unfavored. However, the
introduction of the bounce before the slow roll inflation might
significantly alter this result \cite{Garriga:2012bc}, and also
\cite{Piao:2004me},\cite{Johnson:2011aa}.

In Ref.\cite{Piao:2004me}, it is showed that in different cycle of
cyclic universe, the universe may be in different minimum of a
landscape, in which the bouncing inflation is responsible for the
emergence of observational universe. In Ref.\cite{Sahni:2012er},
it is showed that the inflation after bounce causes the
cosmological hysteresis, which leads to the increase in the
amplitude of cycles.

Thus whether theoretically or observationally, the studying of the
bouncing inflation model is interesting. We will begin with a
clarify of the primordial perturbation in this model in Sec.II. In
Sec.III, we fit the model with the Planck data, and show that this
model may generate the power deficit at low-$l$ and the
hemispherical power asymmetry in CMB, which is consistent with the
Planck data. The Sec.IV is the conclusion. We will briefly
illustrate the model building and discuss the implication of the
result to the eternal inflation scenario in the Appendix.


Note added: While this work is completed,
Ref.\cite{Biswas:2013dry} appeared, in which the authors discussed
the effect of an instantaneous superinflationary phase after the
bounce to the inflationary power spectrum.

\section{The primordial perturbation in bouncing inflation scenario}

We will clarify the results of the primordial perturbation in
bouncing inflation scenario. Here, we require that the bounce
occurs at a higher scale than the inflationary scale and in the
meantime all physical quantities continuously pass through the
bounce. We will see that the result is insensitive with respect to
implementing detail of the bounce.

The quadratic action of the curvature perturbation $\cal R$ is \be
S_2\sim \int d\eta d^3x {a^2M_P^2\epsilon\over c_s^2}\left({{\cal
R}^\prime}^2-{c_s^2}(\partial {\cal R})^2\right), \ee which is
actually universal for single field, e.g. \cite{GM}, where
definition of $\epsilon$ is ${d\over dt}({1/ H})$. The equation of
$\cal R$ in momentum space is \cite{Muk},\cite{KS} \be
u_k^{\prime\prime} +\left(c^2_s k^2-{z^{\prime\prime}\over
z}\right) u_k = 0, \label{uk}\ee  after $u_k \equiv z{\cal R}_k$
is defined, where $'$ is the derivative with respect to conformal
time $\eta=\int dt/a$, $z\equiv a\sqrt{2M_P^2\epsilon}/c_s$. We
have $c_s^2=1$ for canonical scalar field.

When $k^2\simeq z^{\prime\prime}/z$, the perturbation mode is
leaving the horizon. When $k^2\ll z^{\prime\prime}/z$, the
solution of $\cal R$ given by Eq.(\ref{uk}) is \ba {\cal R} & \sim
& C\,\,\,\,\, is\,\,\,{{\rm constant}}\,\,\,{ {\rm
mode}}\label{C}\\ &or &\, D\int {d\eta\over z^2}\,\,\,\,\,
is\,\,\,{{\rm decaying}}\,\,\,{ {\rm mode}} , \label{D}\ea where
the change of $D$ mode is dependent on the evolution of $z$.

Before the bounce the universe is kinetic-dominated, whilst after
the bounce it will get into an inflationary phase, we have
$\epsilon_{\rm inf}\ll 1$, see Appendix for the detailed models.
Thus in conformal time, after adopting an instantaneous matching
between both regimes, we have \ba a & \simeq & a_0\sqrt{1-2{\cal
H}_0\eta}~,
\,\,\,for\,\,{\rm the }\,\,{\rm contraction} \label{leq} \nonumber\\
& & {a_0\over 1-{\cal H}_0\eta}~, \,\,\, for \,\,{\rm the }\,\,{\rm
inflation}. \label{geq} \ea where $\eta<0$ in the contracting phase
and $\eta>0$ in the inflationary, respectively, and $a=a_0$ for
$\eta=0$ is set, ${\cal H}_0$ is the comoving Hubble length at
matching time $\eta=0$, which sets the inflationary energy scale by
$H_{\rm inf}={\cal H}_0/a_0$.

When $k^2\gg {z^{\prime\prime}\over z}$, i.e. the perturbation is
deep inside its horizon, $u_k$ oscillates with a constant
amplitude, \be u_k\sim {1\over \sqrt{2k}} e^{-ik\eta}.
\label{ini}\ee In the contracting phase before inflation, \be
{z^{\prime\prime}\over z}\simeq {-{\cal H}_0^2\over (1-2{\cal H}_0
\eta)^2}, \ee which will increase with time. When $k^2\ll
{z^{\prime\prime}\over z}$, i.e. the perturbation is far outside
the horizon, the solution of Eq.(\ref{uk}) is \be
{u_k}=\sqrt{\pi(1-2{\cal H}_0\eta)\over 8{\cal
H}_0}H_0^{(1)}\left(-k\eta+{k\over 2{\cal H}_0}\right)
 ,\ee where $H_0^1$ is Hankel function of the first kind and
of zeroth order.

In the inflationary phase, \be {z^{\prime\prime}\over z} \simeq
{2{\cal H}_0^2 \over (1-{\cal H}_0\eta)^2} .\ee When $k^2\ll
{z^{\prime\prime}\over z}$, the solution of Eq.(\ref{uk})
is \ba & &{u_k} =\sqrt{-k\eta +{k\over {\cal H}_0}}\nonumber \\
& &\left(C_1 H_{3/2}^{(1)}(-k\eta +{k\over {\cal H}_0})+C_2
H_{3/2}^{(2)}(-k\eta +{k\over {\cal H}_0})\right) \label{vki} ,\ea
where $H_{3/2}^{(1)}$ is Hankel functions of the first kind and of
$3/2$ order, and $H_{3/2}^{(2)}$ is Hankel functions of the second
kind and of $3/2$ order, $C_1$ and $C_2$ are only dependent on
$k$.

When the bounce is nonsingular, all physical quantities should
continuously pass through the bounce. The continuity of curvature
perturbation brings
 \ba C_1 =&
&\sqrt{\pi\over 32{\cal H}_0}e^{-ik\over {\cal H}_0} ((1-{2{\cal
H}_0^2\over k^2}-{2{\cal H}_0\over
k}i)H_0^{(2)}\left({k\over 2{\cal H}_0}\right)\nonumber \\
& & +({{\cal H}_0\over k} +i)H_1^{(2)}\left({k\over 2{\cal
H}_0}\right)) ,\label{c1}\ea \ba C_2 &=& \sqrt{\pi\over 32{\cal
H}_0}e^{ik\over {\cal H}_0} ((1-{2{\cal H}_0^2\over
k^2}+{2{\cal H}_0\over k}i)H_0^{(2)}\left({k\over 2{\cal H}_0}\right)\nonumber\\
& &+({{\cal H}_0\over k} -i)H_1^{(2)}\left({k\over 2{\cal
H}_0}\right)) ,\label{c2}\ea where $H_0^{(2)}$ is Hankel functions
of the second kind and of zeroth order, and $H_1^{(2)}$ is Hankel
functions of the second kind and of first order.

The spectrum of $\cal R$ is \be {\cal P}_{\cal R} = {k^3\over
2\pi^2}\left|{u_k\over z}\right|^2. \label{p}\ee We substitute
Eq.(\ref{vki}) into (\ref{p}), and have the spectrum of curvature
perturbation \ba {\cal P}_{\cal R}& = & {{H}_{\rm inf}^2\over
2\pi^3 M_P^2\epsilon_{\rm inf}}k\left|C_1 -C_2\right|^2
\nonumber\\ & = & {\cal P}_{\cal R}^{\rm inf} {2\over
\pi}k\left|C_1 -C_2\right|^2, \label{ps} \ea \be n_{\cal
R}-1={d\ln{\cal P}_{{\cal R}}\over d\ln k}, \label{nr}\ee where
${\cal P}_{\cal R}^{\rm inf} = {H_{\rm inf}^2\over 4 \pi^2 M_P^2
\epsilon_{\rm inf}}$ is that of the standard slow roll inflation,
which may has a slight red spectrum consistent with the
observation, $C_1$ and $C_2$ are determined by Eqs. (\ref{c1}) and
(\ref{c2}), respectively. Here, we have expanded $H_{3/2}^{(1)}$
and $H_{3/2}^{(2)}$ in term of $-k\eta +{k/{\cal H}_0}\ll 1$, and
used ${\cal H}_0=a_0H_{\rm inf}$

Here, ${\cal H}_0$ is the comoving Hubble length at matching time
$\eta=0$, which implies that ${\cal P}_{\cal R}$ for $k\ll {\cal
H}_0$ is given in the contracting phase, while ${\cal P}_{\cal R}$
for $k \gg {\cal H}_0$ is in the inflationary phase. In $C_1$ and
$C_2$, the Hankel functions $H_{0,1}^{(2)}$ are the function of
$k/{\cal H}_0$. We can expand the Hankel functions in term of
$k\ll {\cal H}_0$ and have approximately \ba {\cal P}_{\cal
R}(k<{\cal H}_0)& \simeq & \frac{H_{\rm inf}^2}{36\pi^4
M_P^2\epsilon_{\rm inf}}(2+\ln\frac{4{\cal
H}_0}{k})^2\frac{k^3}{{\cal H}_0^3} \nonumber\\ & \sim &
\left({k\over {\cal H}_0}\right)^3 \ln\frac{{\cal
H}_0}{k}.\label{P1}\ea Thus on the scale $k\ll {\cal H}_0$, the
spectrum
is strongly blue $\sim k^3$, which is the usual result of PBB
scenario. While for $k \gg {\cal H}_0$, we have \ba {\cal P}_{\cal
R}(k>{\cal H}_0) & \simeq & {H_{\rm inf}^2\over 4 \pi^3 M_P^2
\epsilon_{\rm inf}}(1+\frac{{\cal H}_0}{4k}\sin\frac{2k}{{\cal
H}_0}), \label{P2}\ea which is almost scale invariant but
modulated with a small oscillation. The scale invariance of
spectrum is actually the result of inflationary evolution after
the bounce. The reason is in the contracting phase the
perturbation mode with $k>{\cal H}_0$ is still inside the horizon,
and its evolution is determined by Eq.(\ref{ini}) and is
insensitive to the background at this stage, only when the
corresponding perturbation mode leaves the horizon, which occurs
in the inflationary phase, the perturbation spectrum is determined
by the evolution of the background.

We plot ${\cal P}_{\cal R}$ in (\ref{ps}) as a function of $k$ in
Fig.1. We see that for $k> {\cal H}_0$, the spectrum is almost
scale invariant with a slightly red tilt and an oscillation with a
decaying amplitude, and for $k<{\cal H}_0$ the amplitude of
spectrum decreases rapidly and gets a cutoff, which is consistent
with our analytical results (\ref{P1}) and (\ref{P2}).


\begin{figure}[htbp]
\includegraphics[scale=2,width=7.0cm]{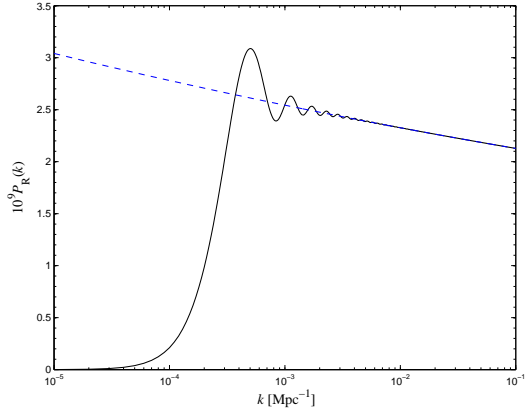}
\caption{Best-fit primordial power spectrum of curvature
perturbations for the pure power law (dashed) and bouncing
inflation (solid) using Planck+WP data.} \label{fig:v1}
\end{figure}

\section{The CMB anomalies with Planck}

\subsection{The power deficit in low-$l$}

\begin{figure}[htbp]
\includegraphics[scale=2,width=7.0cm]{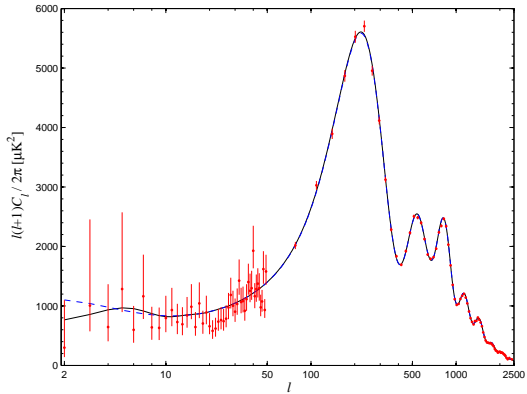}
\caption{Best-fit temperature power spectrum for the pure power law (dashed)
and bouncing inflation (solid) using Planck+WP data. The red points show
the Planck data with 1$\sigma$ errors.}
\label{fig-cls}
\end{figure}

In Eq.(\ref{ps}), ${\cal P}_{\cal R}^{\rm inf}$ may be
parameterized as a power law with \be {\cal P}_{\cal R}^{\rm inf}
= A_{\rm inf} \left(\frac{k}{k_0}\right)^{n_{\rm inf}-1}. \ee Here
we emphasize that in terms of this definition, the spectral index
of curvature perturbation defined in Eq.(\ref{nr}) is $n_{\cal
R}-1 \simeq 3$ for $k\ll {\cal H}_0$ and $n_{\cal R}=n_{\rm inf}$
for $k> {\cal H}_0$. Thus the primordial spectrum (\ref{ps}) is
described by three free parameters, \{$A_{\rm inf}$,$n_{\rm
inf}$,${\cal H}_0$\}. The pivot scale, $k_0$, is chosen to be
$k_0=0.05$Mpc$^{-1}$, roughly in the middle of the logarithmic
range of scales probed by Planck. In addition, cosmological
evolution at late times can be characterized by four free
parameters, \{$\Omega_b h^2,\Omega_c h^2,\Theta_s,\tau$\}, where
$h$ is the dimensionless Hubble parameter such that $H_0 = 100h$
kms$^{-1}$ Mpc$^{-1}$, $\Omega_b h^2$ and $\Omega_c h^2$ are the
physical baryon and dark matter densities relative to the critical
density, $\Theta_s$ is the ratio of the sound horizon to the
angular diameter distance at the photon decoupling, and $\tau$ is
the reionization optical depth. We impose a uniform prior on the
logarithm of ${\cal H}_0$ in the range $[-12,-4]$. For the other
parameters, prior ranges are chosen to be much larger than the
posterior. In order to compute the theoretical CMB power spectrum,
we modify the Boltzmann CAMB code in~\cite{lew99}. In
Fig.~\ref{fig-cls} we plot the angular power spectrum for the pure
power law (dashed) and bouncing inflation with the best-fit value
of $\ln ({\cal H}_0/{\rm Mpc}^{-1})=-8.60$ (solid). Compared to
the standard power-law model, the $C_l$ spectrum in the bouncing
Universe is suppressed in the quadrupole and octupole. Moreover, a
small bump around $l=6$ arises from the oscillations of the
primordial power spectrum at large scales.

We use the combination of the Planck CMB temperature power
spectrum~\cite{Ade:2013xsa,Ade:2013uln} with the WMAP large-scale
polarization data~\cite{pag07} (denoted ``Planck+WP''). The Planck
temperature likelihood is based on a hybrid approach, which
combines a pixel-based likelihood at low multipoles ($2\le l \le
49$) with a Gaussian likelihood approximation at high multipoles
($50 \le l \le 2500$). The Planck high-$l$ likelihood involves 14
nuisance parameters to describe unresolved small-scale foreground
and CMB secondary anisotropies. Since Planck doesn't release
polarization data, the WMAP polarization data at low multipoles
($2\le l \le 23$) is used to constrain the optical depth.

\begin{figure}[htbp]
\includegraphics[scale=2,width=14.0cm]{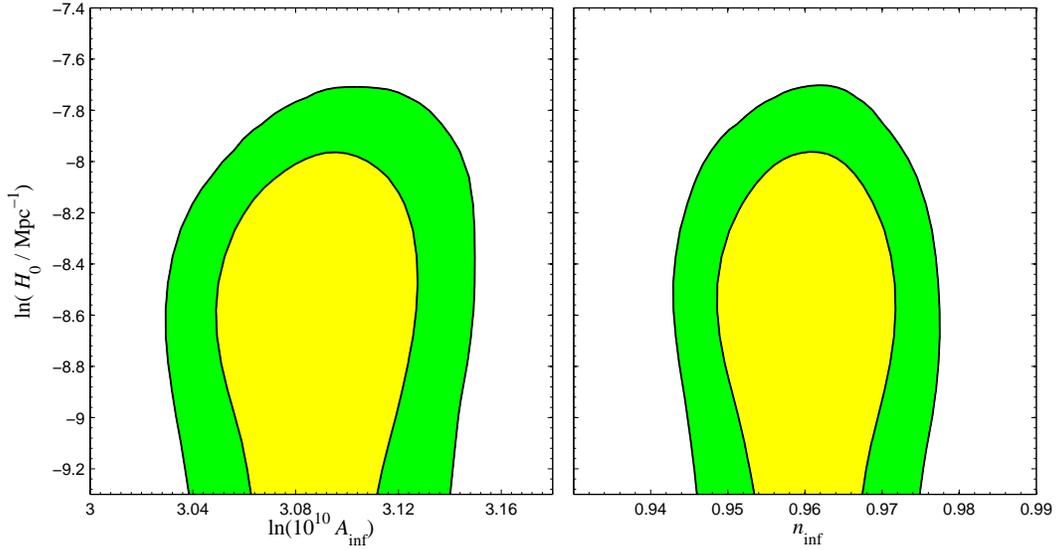}
\caption{Two-dimensional joint marginalized constraints (68\% and 95\% confidence level)
on ${\cal H}_0$, $A_{\rm inf}$ and $n_{\rm inf}$, derived from the Planck+WP data.}
\label{fig-contour}
\end{figure}

\begin{figure}[htbp]
\includegraphics[scale=2,width=7.0cm]{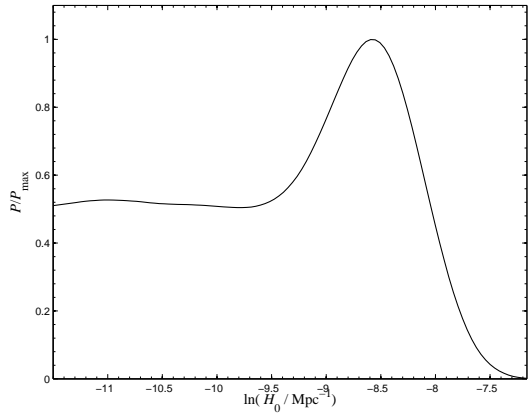}
\caption{Marginalized posterior distributions for ${\cal H}_0$ from the Planck+WP data.}
\label{fig-dist}
\end{figure}

In our analysis we use a modified version of the publicly
available CosmoMC package to explore the parameter space by means
of Monte Carlo Markov chains technique~\cite{lew02}. From the
Planck+WP data we find the best-fit values of $\ln ({\cal
H}_0/{\rm Mpc}^{-1})=-8.60$, $\ln (10^{10} A_{\rm inf})=3.084$ and
$n_{\rm inf}=0.961$ with $-\ln({\cal L}_{\rm max})=4901.6$. This
indicates that the bouncing inflation model can improve the fit to
the data with $\Delta\chi_{\rm eff}^2\approx -4.6$ with respect to
the standard power-law model. However, a phenomenological
exponential-form cutoff of the primordial power spectrum improves
the fit only with $\Delta\chi_{\rm eff}^2\approx -2.9$ reported
in~\cite{Ade:2013uln}. Moreover, the exponential-form cutoff
in~\cite{Ade:2013uln} is described by two parameters, the cutoff
steepness $\lambda_c$ and the cutoff scale $k_c$. In the bouncing
inflation model, the cutoff is characterized by only one parameter
${\cal H}_0$. We show the joint constraints on ${\cal H}_0$,
$A_{\rm inf}$ and $n_{\rm inf}$ in Fig.~\ref{fig-contour}. Since
${\cal H}_0$ characterizes local features in the power spectrum
while $A_{\rm inf}$ and $n_{\rm inf}$ characterize the global
shape of the power spectrum (see~\cite{guo11} for a general shape
reconstructed from CMB data), there is nearly no correlation
between them, as shown in Fig.~\ref{fig-contour}. The marginalized
posterior distribution of ${\cal H}_0$ is shown in
Fig.~\ref{fig-dist}, which illustrates the asymmetric shape of the
likelihood functions.

\subsection{The hemispherical power asymmetry}

Recently, the Planck collaboration has reported a hemispherical
power asymmetry in CMB \cite{Ade:2013nlj}, which conformed a
similar result of WMAP \cite{Eriksen:2007pc}, but has better
precision. It could be thought that this power asymmetry might
result from a superhorizon perturbation crossing the observable
universe. We will estimate the hemispherical power asymmetry in
the bouncing inflation, along the line of
Ref.\cite{Erickcek:2008sm} by Erickcek et.al and
Ref.\cite{Lyth:2013vha} by Lyth.

The CMB power asymmetry might be modeled as a dipole modulation of
the power
\cite{Prunet:2004zy},\cite{Gordon:2005ai},\cite{Gordon:2006ag}.
This modulation can be explained in light of the spatial change of
the power spectrum of primordial curvature perturbation ${\cal R}$, \be {\cal P}^{1/2}_{\cal
R}(k,\mathbf{x})=\left(1+ A(k){\hat{\mathbf{p}}\cdot
\mathbf{x}\over x_{\rm ls}}\right) {\cal P}^{1/2}_{{\cal R}}(k), \ee
where $A(k)$ is the amplitude of the modulation,
the unit vector $\hat{\mathbf{p}}$ is the dipole modulation direction, $x_{\rm ls}$ is the
distance to the last scattering surface, and ${\cal
P}^{1/2}_{{\cal R}}(k)$ is given by Eq.(\ref{ps}).

The asymmetry $A(k)$ can be calculated as, \ba A(k) & = &
{|\nabla{\cal P}^{1/2}_{{\cal R}}(k,\mathbf{x})|\over {\cal
P}^{1/2}_{{\cal R}}}\,x_{\rm ls}\nonumber\\ & = & \left(
{d\ln{{\cal
P}^{1/2}_{{\cal R}}}\over d\ln{k}}\left|\nabla\ln{k}\right|\right) x_{\rm ls} \nonumber\\
& = & \left(1-\epsilon_{Per}\right)\left[{n_{\cal R}(k)-1\over
2}\right]k_{\rm L} x_{\rm ls}{\cal P}^{1/2}_{{\cal R},{\rm L}}\,,
\label{tp}\ea where ${\cal P}_{{\cal R},{\rm L}}$ is the amplitude
of the power spectrum of a single modulating mode $k_{\rm L}$,
i.e.,
\be {\cal P}_{{\cal R}}(k)={\cal P}_{{\cal R},{\rm L}}
\,\,\delta(\ln{k}-\ln{k_{\rm L}}),\label{kl}\ee and for this
modulating mode we have $|\nabla {\cal R}_{\rm L}|=k_{\rm L}{\cal
R}_{\rm L}=k_{\rm L}{\cal P}^{1/2}_{{\cal R},{\rm L}}$.
Here,
$\epsilon_{Per}=\epsilon_{C}$ for the contracting phase and
$\epsilon_{Per}=\epsilon_{\rm inf}$ for the inflationary phase.
The inflationary result is recovered for $\epsilon_{Per}\ll 1$
\cite{Erickcek:2008sm},\cite{Lyth:2013vha}. Here,
$1-\epsilon_{Per}$ arises from the dependence of $H$ on the time in
Eq.(\ref{nr}), since $k=aH$.

The maximum achievable value of $A(k)$ is given by the limits on
the terms on the right-hand side of Eq.(\ref{tp}). The
perturbation amplitude of the superhorizon mode with $k_{\rm L}$ can be constrained by its
contribution to the CMB quadrupole, i.e., the Grishchuk-Zel'dovich
effect. This has been computed using the Sachs-Wolfe approximation in Ref.\cite{Lyth:2013vha} as
\ba
C_2^{\rm GZ} & = & {4\pi\over 25}\int^{1/x_{\rm ls}}_0{dk\over
k}\left({(k x_{\rm ls})^2\over 15}\right)^2 {\cal P}_{{\cal
R}}(k)\nonumber\\ & = & {4\pi\over 25}\left({(k_L x_{\rm ls})^2\over
15}\right)^2 {\cal P}_{{\cal R},{\rm L}}\,. \label{gze} \ea
In deriving the second line of (\ref{gze}) we have used Eq.(\ref{kl}).
Requiring $\sqrt{C_2^{\rm GZ}}$ to be $\lesssim 3
\times$ the measured rms value of the quadrupole, we have \be (k_L
x_{\rm ls})^4 {\cal P}_{{\cal R},{\rm L}} \lesssim 16\,\times \,
10^{-8}, \ee which gives $(k_L x_{\rm ls}){\cal P}^{1/2}_{{\cal R},{\rm L}}\lesssim 0.02$, since ${\cal
P}_{{\cal R},{\rm L}}\lesssim 1$ for the perturbation theory to apply.
Plugging this result into Eq.(\ref{tp}), we have an upper bound
on the modulation amplitude of the CMB power asymmetry, \be |A(k)| \lesssim
0.02\,\left|{n_{\cal R}(k)-1\over
2}\right|\left|1-\epsilon_{Per}\right|. \ee
This amplitude is measured to be $|A|=0.07\pm 0.02$ from the 5-year WMAP analyses~\cite{Hoftuft:2009rq},
which is consistent with the Planck results~\cite{Ade:2013nlj}.

In single field inflationary scenario, $\epsilon_{Per}\ll 1$ and
$n_{\rm inf}-1\sim 0.04$. Thus we have $|A(k)|\sim 10^{-4}$, which
is too small to fit the observation, as pointed out in
Refs. \cite{Erickcek:2008sm},\cite{Lyth:2013vha}. However, the case is
altered in curvaton scenario, see
\cite{Lyth:2013vha},\cite{Wang:2013lda}.

In bouncing inflation scenario discussed here, on large angular
scale $1/k>1/{\cal H}_0$, the curvature perturbation origins from
the fluctuation of $\phi$ during the contraction. We have $n_{\cal
R}-1\simeq 3$ and $\epsilon_{Per}\sim 3$, as have been calculated
in Sec.II. Thus in this scenario the power asymmetry on large
angular scale is \be |A_{\rm B}(k)|\lesssim 0.06, \label{Ak}\ee
which is consistent with Planck data. The power spectrum at
intermediate and small angular scales is that of slow roll
inflation, thus the corresponding power asymmetry is small, which
is consistent with the constraint from the SDSS sample of quasars
\cite{Hirata}.

Here, (\ref{Ak}) applies only to scales $1/k \gtrsim 1/{\cal
H}_0$, while the required range is $x_{\rm ls}/60 \lesssim 1/k \ll
x_{\rm ls}$ \footnote{We thank David H. Lyth for pointing out this
to us.}. Our best-fit value is $1/{\cal H}_0\simeq 5{\rm Gpc}$,
which corresponds to $1/{\cal H}_0\simeq x_{\rm ls}/3$, since the
distance to the last scattering surface is estimated as $x_{\rm
ls}\simeq 14{\rm Gpc}$. This result seems to imply that our model
has a tension with the observation. However, as shown in Fig.3, at
$2\sigma$ confidence level ${\cal H}_0$ may be $1/{\cal H}_0=
1.5{\rm Gpc} \simeq x_{\rm ls}/9$, and further, at $3\sigma$ which
though was not plotted, it may be $1/{\cal H}_0\simeq x_{\rm
ls}/30$. Thus our model is consistent with Planck's constraints at
$3\sigma$. The observation has placed strong constraints on our
model, which makes it easily falsified by further Planck data.

\section{Conclusion}

Recently, the Planck collaboration has released the data of the
power on cosmic microwave background, which is consistent
with the slow roll inflationary model. However, the Planck data
also show a power deficit at $l\lesssim 40$ and a hemispherical
power asymmetry in CMB, which conformed the early observations by
WMAP. This result is intriguing, since it might be a hint of the
physics at the epochs before the inflation.

We continue studying the bouncing inflation model. In this model,
we assume that initially the universe is in a contracting phase,
and after the bounce it begins to inflate. The contraction before
the bounce leads to that the primordial power spectrum on large
angular scales $1/k>1/{\cal H}_0$ has the spectral index $n_{\cal
R}-1\simeq 3$, where ${\cal H}_0$ is a new degree of freedom,
which sets the cutoff scale.
We find that this spectrum generates not only the power deficit at
low-l, but also the hemispherical power asymmetry in CMB, which
may be consistent with the Planck data. Thus our model can explain
the CMB anomalies, which may be falsified by further Planck data.

The bouncing inflation model not only sets a natural initial
condition for the beginning of the slow roll inflation,
significantly but also is connected with the preinflationary
physics. We discussed the model building in the Appendix. It is
interesting to embed a bouncing model into a fundamental theory.
In principle, depending on the implemented detail of the bounce,
the models may be different. However, these details do not
quanlitatively affect the result of the primordial spectrum given
here.

We also showed that in the eternal inflation scenario, the
bouncing inflation might be a favored channel to implement the
slow roll inflation. Thus it is interesting to have a detailed
study in a string landscape motivated well.

\textbf{Acknowledgments}

We thank David H. Lyth for helpful discussions. ZKG is supported
by the project of Knowledge Innovation Program of Chinese Academy
of Science, NSFC under Grant No.11175225, and National Basic
Research Program of China under Grant No.2010CB832805. YSP is
supported by NSFC under Grant No.11075205, 11222546, and National
Basic Research Program of China, No.2010CB832804. We used CosmoMC
and CAMB. We acknowledge the use of the Planck data and the Lenovo
DeepComp 7000 supercomputer in SCCAS.

\section*{Appendix A: The models of bouncing inflation}

In this Appendix, we will discuss some models of the bouncing
inflation.

\begin{figure}[htbp]
\includegraphics[scale=2,width=10.0cm]{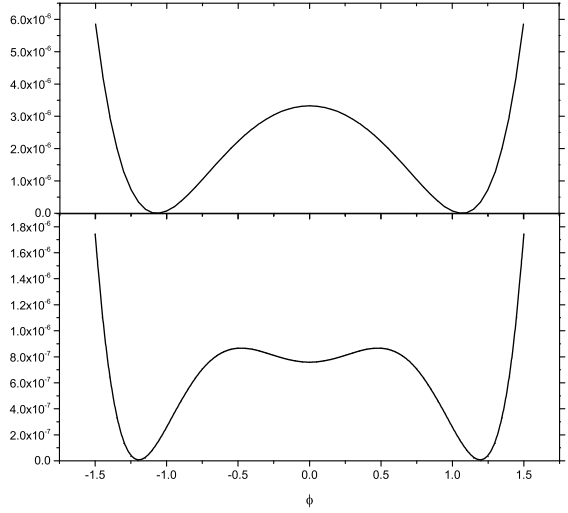}
\caption{In the upper panel, we choose the parameter
$\lambda_2=-1$, $M_\phi=3.0\times10^{-3}$,
$\lambda_4=1.0\times10^{-6}$,
$\lambda_6=1.0\times10^{-6}$,$\lambda_0=3.3\times10^{-6}$. In the
lower panel, we choose the parameter
$\lambda_2=1$,$M_\phi=\sqrt{2}\times10^{-3}$,
$\lambda_4=-1.0\times10^{-5}$,
$\lambda_6=1.0\times10^{-3}$,$\lambda_0=7.6\times10^{-5}$ }
\label{fig:v}
\end{figure}

The Lagrangian is \ba
\mathcal{L}=\frac{1}{2}\partial_{\mu}\phi\partial^{\mu}\phi-\left(\frac{1}{2}\partial_{\mu}\psi\partial^{\mu}\psi\right)^{2/3}-V(\phi),
\ea where the potential is only the function of the field $\phi$.
Here, we regard the potential as \ba V={1\over
2}M_{\phi}^2\lambda_2\phi^2+{1\over 4}\lambda_{4}\phi^4+\lambda_6
{\phi^6\over M_P^2}+\lambda_0. \ea This potential is a Higgslike
potential for the parameters $\lambda_2<0$, $\lambda_4>0$ and
$\lambda_6=0$. While for $\lambda_4<0$ and
$\lambda_2,\lambda_6>0$, this potential corresponds to that from
the minimal supersymmetric standard model
e.g.\cite{Allahverdi:2006we}. There might be two or three minima
in this potential, dependent of the values of parameters, see
Fig.\ref{fig:v}.

Here, $\psi$ is the ghost field, whose only role is to simply
implement the bounce. In principle, the ghost instability may be
dispelled by applying the Galileon interaction \cite{Qiu1}. The
bounce may be also implemented like in e.g.\cite{GV2},\cite{LWC}
for PBB scenario, \cite{Khoury:2001wf} for ekpyrotic scenario.
The Lagrangian of $\psi$ is specially selected for convenience,
since it may lead to an analytical solution of $a$ around the
bounce.

\begin{figure}[htbp]
\includegraphics[scale=2,width=12.0cm]{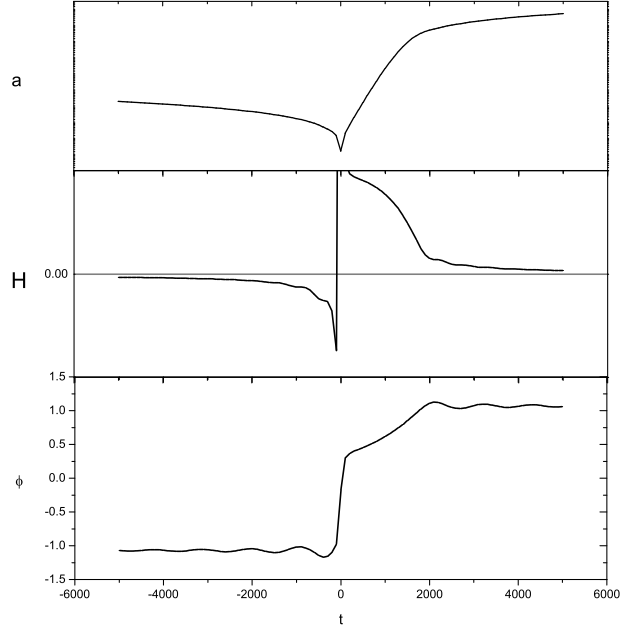}
\caption{The evolutions of $\phi$, $H$ and $a$ with the time for
the potential in the upper panel of Fig.\ref{fig:v}}
\label{fig:pah}
\end{figure}

\begin{figure}[htbp]
\includegraphics[scale=2,width=7.0cm]{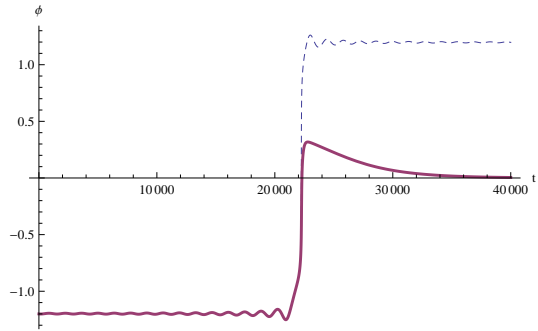}
\caption{The evolution of $\phi$ for the potential in the lower
panel of Fig.\ref{fig:v}. The solid line corresponds the field
$\phi$ rolls from the left minimum to the middle minimum and the
dashed line corresponds the field rolls from the left minimum to
the right minimum.}\label{fig:phi}
\end{figure}

\begin{figure}[htbp]
\includegraphics[scale=2,width=7.0cm]{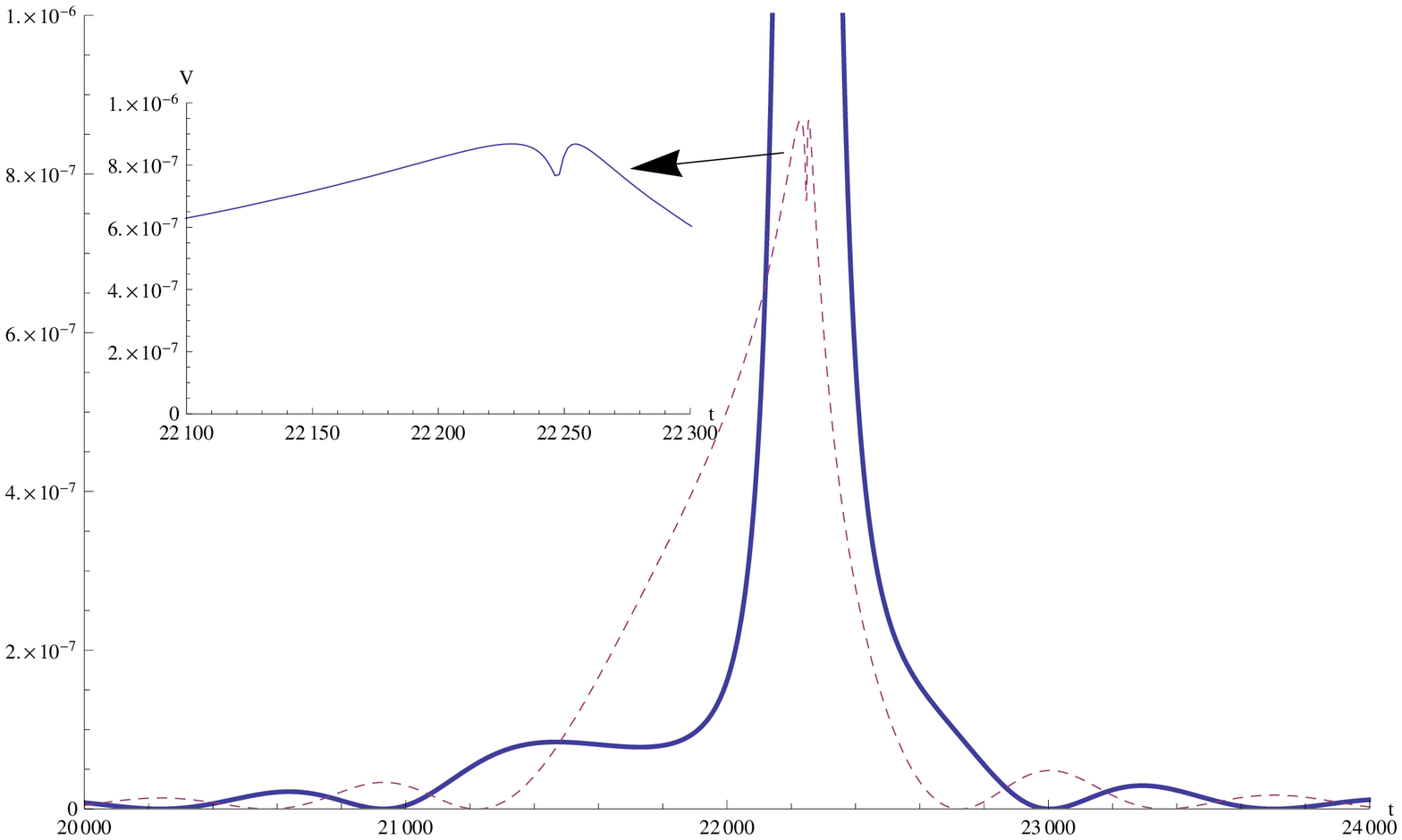}
\caption{The solid line is the evolution of kinetic energy for the
potential in the lower panel of Fig.\ref{fig:v}, while the dashed
line is the evolution of potential energy. }\label{fig:tv1}
\end{figure}

We plot the evolution of $\phi$, $H$ and $a$ in Fig.\ref{fig:pah}
for the potential in the upper panel of Fig.\ref{fig:v}, and the
evolutions of $\phi$, the kinetic energy and the potential energy
in Figs.\ref{fig:phi} and \ref{fig:tv1} for the potential in the
lower panel of Fig.\ref{fig:v}. The universe initially is in a
contracting phase, and the field $\phi$ is in one among the minima
of its potential. We see that before the bounce, the field will
climb up along its potential
\cite{Piao:2003zm},\cite{Kanekar:2001qd}, and its kinetic energy
${\dot\phi}^2$ will become dominated, while after the bounce, the
kinetic energy of $\phi$ will be rapidly diluted and the universe
will get into an inflationary phase, and finally the field will
roll down along the potential to the other minima. Here, the
contracting phase actually provides a homogeneous patch for the
beginning of slow roll inflation, which helps to relax the initial
conditions problem argued in Ref.\cite{Ijjas:2013vea}.

The Lagrangian of $\psi$ implies $\rho_{\psi}=c_{\psi}/a^{12}$.
When ${\dot\phi}^2$ is dominated, we have $\rho_\phi=c_{\phi}/a^6$
for $\phi$ field. Thus the Friedmann equation is \be \int dt=\int
\frac{da^6/6M_P\sqrt{3}}{\sqrt{c_{\phi}a^6-{c_\psi}}}.\ee Thus we
have \be \int dt=\sqrt{a^6-{c_\psi\over c_\phi}}/\sqrt{3
c_{\phi}}M_P.
\ee There is a bounce at $a^6_B={c_\psi\over c_\phi}$, while $a\sim
t^{1/3}$ when $a$ deviates from $a_B$, which is consistent with
Fig.\ref{fig:pah}.

The field $\phi$ will walk with certain distance during
${\dot\phi}^2$ is dominated, before it finally lands in an
inflationary region. Here, ``land" means that the effective
potential of field begins to become dominated. The change of
$\phi$ before the field lands is $\Delta\phi$. We have
$M_P^2H^2={\dot\phi}^2/6$, which gives
\cite{Sahni:2012er},\cite{Piao:2004hr} \be {\Delta\phi} \sim
M_P\ln{\left({H_B\over H_{kin}}\right)}, \ee where $H=1/(3t)$ for
$a\sim t^{1/3}$ is applied, and $H_B$ is the Hubble parameter
before and after the bounce and $H_{kin}$ is that at the time when
the kinetic energy of field begins to dominate. We generally have
$H_B
> H_{kin}$, which implies $\Delta\phi\gtrsim M_P$.

\section*{Appendix B: The implication for the eternal inflation scenario}

\begin{figure}[htbp]
\includegraphics[scale=2,width=7.0cm]{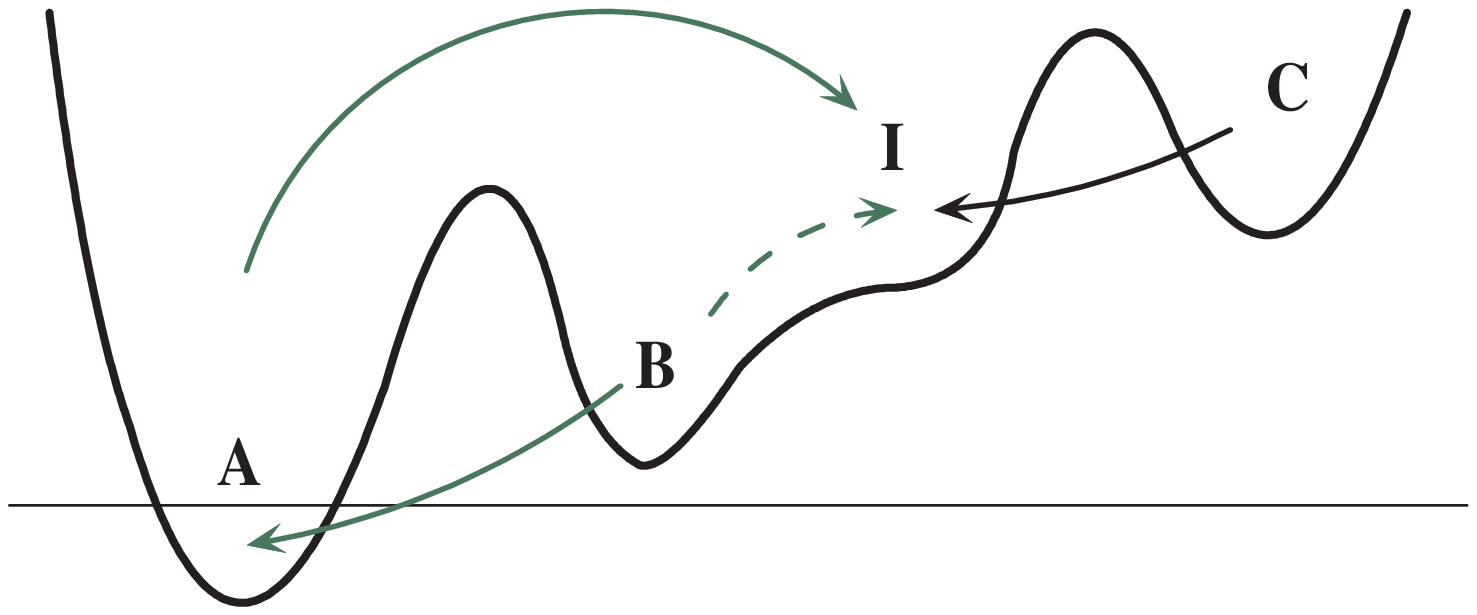}
\caption{A landscape of effective potential. We are interested in
the ratio of probabilities of the different channels to the slow
roll inflationary region, which are plotted with the black line,
blue solid line and dashed line, respectively.}\label{fig:bounce}
\end{figure}

In the eternal inflation scenario \cite{V1}, an infinite number of
universes will be spawned in the eternally inflating background.
It might be thought that a phase of the slow roll inflation and
reheating is required for a spawned universe becoming our
observable universe.

The slow roll inflation should start in a high scale, which is
required to insure that the amplitude of primordial perturbation
is consistent with the observations and the reheating temperature
is suitable for a hot big bang model. In this sense, if the scale
of the eternally inflating background is very low, the spawning of
observational universe will requires a large uptunneling, which is
exponentially unfavored.

However, the introduction of the nonsingular bounce might
significantly alter this result
\cite{Garriga:2012bc},\cite{Piao:2004me},\cite{Johnson:2011aa}.
Here, we will briefly revisit this issue with the bouncing
inflation. We will show that the bouncing inflation is a favored
channel to the slow roll inflation in a given landscape.

In a landscape of the effective potential in Fig.\ref{fig:bounce},
we have a AdS minimum `A', a dS minimum `B' with lower energy, a
dS minimum with higher energy and a slow-roll inflationary region
`I'. The transitions in this landscape are `B' $\rightarrow$ `A',
`B' $\rightarrow$ `I' and `I' $\leftrightharpoons$ `C'. In
addition, `I' may also classically roll into `B', and the AdS
crunch in `A' is replaced with the bounce, and the corresponding
probabilities of bouncing to `B', `C' and `I' are $Q_B$, $Q_C$ and
$Q_I$, respectively, with $\sum_i Q_i = 1$.

We follow Ref.\cite{GV}. The rate equation describing the fractions
$f_j$ in corresponding regions is $ {d\mathbf{f}\over
dt}=\mathbf{\cal M}\mathbf{f}$, where \be \mathbf{f}=
\left(\begin{array}{c}
f_A\\
f_B\\
f_I\\
f_C
\end{array}\right), \ee  \be \mathbf{\cal M} = \left(\begin{array}{cccc}
-1 & \kappa_{AB} & \kappa_{AI} & 0 \\
Q_{BA} & -\kappa_{AB}-\kappa_{IB} & S_{BI} & 0 \\
Q_{IA}  & \kappa_{IB} & -\kappa_{AI}-S_{BI}-\kappa_{CI} & \kappa_{IC}\\
Q_{CA} & 0 & \kappa_{CI} & -\kappa_{IC}
\end{array}\right), \label{matrix}\ee where
$\kappa_{ij}=4\pi\Gamma_{ij}/3H_j^3$ is the transition rate, and
$\Gamma_{ij}$ is the nucleation rate of bubble, and $S_{BI}\sim
1/t_{BI}$, $t_{BI}={\cal N}/H_{\rm inf}$ is the time the slow roll
inflation lasts and ${\cal N}$ is the efolds number. The
distributions $f_j$ will be fixed at late time. Thus we have \ba
{f_B} & = & {1+Q_{BA}{\kappa_{AI}\over S_{BI}}\over
\kappa_{AB}+(\kappa_{AB}+\kappa_{IB}){\kappa_{AI}\over
S_{BI}}}f_A\nonumber\\ &\simeq & {1\over
\kappa_{AB}(1+{\kappa_{AI}\over S_{BI}})}f_A, \label{BA}\\
f_C & \simeq & {Q_{CA}+{\kappa_{CI}(1-Q_{BA})\over
S_{BI}(1+\kappa_{AI}/S_{BI})}\over \kappa_{IC}} f_A \nonumber\\ &
\simeq & {Q_{CA}\over \kappa_{IC}}f_A, \label{CA}\ea where
$Q_{BA}{\kappa_{AI}\over S_{BI}}\ll 1$ and $\kappa_{IB}\ll
\kappa_{AB}$ are used. We have \be {f_C\over f_B}\simeq
{Q_{CA}\kappa_{AB}\over \kappa_{IC}}, \ee which is consistent with
the result of Garriga and Vilenkin \cite{Garriga:2012bc}, i.e. the
ratio is not suppressed by the small uptunnelling rate.

Here, we are interested in the ratio of probabilities of the
different channels to the slow-roll inflationary region. Here, one
channel is the AdS bounce from `A', the others are the
uptunnelling from `B' and the tunnelling from `C'. We have, after
noting the corresponding terms with a plus sign at the right side
of the $f_I$ equation in Eqs.(\ref{matrix}),
\ba{\dot {\cal P}}_{\rm Ainf} & = & Q_{IA}f_A,\\ {\dot {\cal
P}}_{\rm Binf} & = & \kappa_{IB}f_B,\\ {\dot {\cal P}}_{\rm Cinf}
& = & Q_{IC}f_C \ea for these channels, respectively, in which
${\dot {\cal P}}$ denotes the incoming probability current into
the slow roll inflationary region, as defined in
\cite{Linde:2006nw}, and the subscript ``Ainf" denotes that from
`A' into the inflationary region. Thus the ratio of ${\cal P}_{\rm
Ainf}$ to $ {\cal P}_{\rm Binf}$ is given by \ba {{\cal P}_{\rm
Ainf}\over {\cal P}_{\rm Binf}} = {Q_{IA}\int f_Adt\over
\kappa_{IB}\int f_B dt}\sim {Q_{IA}\kappa_{AB}\over
\kappa_{IB}}(1+ {\kappa_{AI}\over S_{BI}})\gg 1, \label{PB}\ea
where we have made the integral for both sides of Eq.(\ref{BA}),
and substituted it into this equation. We generally have
$\kappa_{AB}\gg \kappa_{IB}$, since $\kappa_{IB}$ is that of the
uptunnelling. Thus Eq.(\ref{PB}) implies that, compared with the
channel of uptunnelling to slow roll inflation, the bouncing
inflation is favored exponentially.

The ratio of ${\cal P}_{\rm Ainf}$ to $ {\cal P}_{\rm Cinf}$ is
given similarly by \ba {{\cal P}_{\rm Ainf}\over {\cal P}_{\rm
Cinf}}   = {Q_{IA}\int f_Adt\over \kappa_{IC}\int f_C dt} \sim
{Q_{IA}\over Q_{CA}}\sim 1. \label{PC}\ea Thus in a given
landscape, the bouncing inflation and the inflationary bubble from
`C' have almost equal possibility. However, it should be noticed
that in Eq.(\ref{CA}), if $Q_{CA}$ is negligible, we will have
${{\cal P}_{\rm Ainf}/{\cal P}_{\rm Cinf}}\gg 1$, in which
$Q_{CA}$ is the contribution from the AdS bounce.



\begin{thebibliography}{99}

\bibitem{Ade:2013xsa}
  P.~A.~R.~Ade {\it et al.}  [ Planck Collaboration],
  arXiv:1303.5062 [astro-ph.CO].

\bibitem{Ade:2013uln}
  P.~A.~R.~Ade {\it et al.}  [Planck Collaboration],
  arXiv:1303.5082 [astro-ph.CO].

\bibitem{Ade:2013nlj}
  P.~A.~R.~Ade {\it et al.}  [Planck Collaboration],
  arXiv:1303.5083 [astro-ph.CO].

\bibitem{Eriksen:2007pc}
  H.~K.~Eriksen, A.~J.~Banday, K.~M.~Gorski, F.~K.~Hansen and P.~B.~Lilje,
  Astrophys.\ J.\  {\bf 660}, L81 (2007)  [astro-ph/0701089].

\bibitem{Hoftuft:2009rq}
  J.~Hoftuft, H.~K.~Eriksen, A.~J.~Banday, K.~M.~Gorski, F.~K.~Hansen and P.~B.~Lilje,
  Astrophys.\ J.\  {\bf 699}, 985 (2009)  [arXiv:0903.1229 [astro-ph.CO]].

\bibitem{Erickcek:2008sm}
  A.~L.~Erickcek, M.~Kamionkowski and S.~M.~Carroll,
  Phys.\ Rev.\ D {\bf 78}, 123520 (2008)  [arXiv:0806.0377
  [astro-ph]].

\bibitem{Lyth:2013vha}
  D.~H.~Lyth,
  arXiv:1304.1270 [astro-ph.CO].

\bibitem{Wang:2013lda}
  L.~Wang and A.~Mazumdar,
  arXiv:1304.6399 [astro-ph.CO].

\bibitem{McDonald:2013aca}
  J.~McDonald,
  arXiv:1305.0525 [astro-ph.CO].

\bibitem{Namjoo:2013fka}
  M.~H.~Namjoo, S.~Baghram and H.~Firouzjahi,
  arXiv:1305.0813 [astro-ph.CO].

\bibitem{Contaldi:2003zv}
  C.~R.~Contaldi, M.~Peloso, L.~Kofman and A.~D.~Linde,
  JCAP {\bf 0307}, 002 (2003)  [astro-ph/0303636];
  G.~Nicholson and C.~R.~Contaldi,
  JCAP {\bf 0801}, 002 (2008)  [astro-ph/0701783].


\bibitem{Piao:2003zm}
  Y.~-S.~Piao, B.~Feng and X.~-m.~Zhang,
  Phys.\ Rev.\ D {\bf 69}, 103520 (2004)  [hep-th/0310206];
  Y.~-S.~Piao,
  Phys.\ Rev.\ D {\bf 71}, 087301 (2005)  [astro-ph/0502343].

\bibitem{Piao:2003hh}
  Y.~-S.~Piao, S.~Tsujikawa and X.~-m.~Zhang,
  Class.\ Quant.\ Grav.\  {\bf 21}, 4455 (2004)  [hep-th/0312139].

\bibitem{Powell:2006yg}
  B.~A.~Powell and W.~H.~Kinney,
  Phys.\ Rev.\ D {\bf 76}, 063512 (2007)  [astro-ph/0612006].



\bibitem{Boyanovsky:2006qi}
  D.~Boyanovsky, H.~J.~de Vega and N.~G.~Sanchez,
  Phys.\ Rev.\ D {\bf 74}, 123006 (2006)  [astro-ph/0607508];
  D.~Boyanovsky, H.~J.~de Vega and N.~G.~Sanchez,
  Phys.\ Rev.\ D {\bf 74}, 123007 (2006)  [astro-ph/0607487];
  \bibitem{Destri:2008fj}
  C.~Destri, H.~J.~de Vega and N.~G.~Sanchez,
  Phys.\ Rev.\ D {\bf 78}, 023013 (2008)  [arXiv:0804.2387
  [astro-ph]];
  D.~Boyanovsky, C.~Destri, H.~J.~De Vega and N.~G.~Sanchez,
  Int.\ J.\ Mod.\ Phys.\ A {\bf 24}, 3669 (2009)  [arXiv:0901.0549 [astro-ph.CO]].


\bibitem{Mielczarek:2008pf}
  J.~Mielczarek,
  JCAP {\bf 0811}, 011 (2008)  [arXiv:0807.0712 [gr-qc]];
  J.~Mielczarek, M.~Kamionka, A.~Kurek and M.~Szydlowski,
  JCAP {\bf 1007}, 004 (2010)  [arXiv:1005.0814 [gr-qc]].

\bibitem{Mortonson:2009xk}
  M.~J.~Mortonson and W.~Hu,
  Phys.\ Rev.\ D {\bf 80}, 027301 (2009)  [arXiv:0906.3016 [astro-ph.CO]].


\bibitem{Liu:2010fm}
  J.~Liu, Y.~-F.~Cai and H.~Li,
  J.\ Theor.\ Phys.\  {\bf 1}, 1 (2012)  [arXiv:1009.3372 [astro-ph.CO]].

\bibitem{Dudas:2012vv}
  E.~Dudas, N.~Kitazawa, S.~P.~Patil and A.~Sagnotti,
  JCAP {\bf 1205}, 012 (2012)  [arXiv:1202.6630 [hep-th]].

\bibitem{BouhmadiLopez:2012by}
  M.~Bouhmadi-Lopez, P.~Chen, Y.~-C.~Huang and Y.~-H.~Lin,
  arXiv:1212.2641 [astro-ph.CO].







\bibitem{MGV} M. Gasperini and G. Veneziano, Astropart. Phys.
\textbf{1} 317 (1993).

\bibitem{Khoury:2001wf}
  J.~Khoury, B.~A.~Ovrut, P.~J.~Steinhardt and N.~Turok,
  Phys.\ Rev.\ D {\bf 64}, 123522 (2001)  [hep-th/0103239];
  E.~I.~Buchbinder, J.~Khoury and B.~A.~Ovrut,
  Phys.\ Rev.\ D {\bf 76}, 123503 (2007)  [hep-th/0702154];


\bibitem{GV2} M. Gasperini, G. Veneziano, Phys. Rept. \textbf{373}, 1
(2003).

\bibitem{LWC} J.E. Lidsey, D. Wands and E.J. Copeland, Phys. Rept. \textbf{337},
343 (2003).








\bibitem{Garriga:2012bc}
  J.~Garriga and A.~Vilenkin,
  arXiv:1210.7540 [hep-th];
  A.~Vilenkin,
  AIP Conf.\ Proc.\  {\bf 1514}, 7 (2012)  [arXiv:1301.0121
  [hep-th]].

\bibitem{Piao:2004me}
  Y.~-S.~Piao,
  Phys.\ Rev.\ D {\bf 70}, 101302 (2004)  [hep-th/0407258]; Y.S. Piao, Phys. Lett. \textbf{B677}, 1 (2009); Phys. Lett. \textbf{B691},
225 (2010).


\bibitem{Johnson:2011aa}
  M.~C.~Johnson and J.~-L.~Lehners,
  Phys.\ Rev.\ D {\bf 85}, 103509 (2012)  [arXiv:1112.3360
  [hep-th]];
  J.~-L.~Lehners,
  Phys.\ Rev.\ D {\bf 86}, 043518 (2012)  [arXiv:1206.1081
  [hep-th]].

\bibitem{Sahni:2012er}
  V.~Sahni and A.~Toporensky,
  Phys.\ Rev.\ D {\bf 85}, 123542 (2012)  [arXiv:1203.0395
  [gr-qc]].

\bibitem{Biswas:2013dry}
  T.~Biswas and A.~Mazumdar,
  arXiv:1304.3648 [hep-th].


\bibitem{GM} J. Garriga, V.F. Mukhanov, Phys. Lett. \textbf{B458}, 219
(1999).

\bibitem{Muk} V.F. Mukhanov, JETP lett. 41, 493 (1985); Sov. Phys. JETP. 68,
1297 (1988).

\bibitem{KS} H. Kodama, M. Sasaki, Prog. Theor. Phys.
Suppl. 78 1 (1984).







\bibitem{lew99}
  A.~Lewis, A.~Challinor and A.~Lasenby,
  Astrophys.\ J.\ {\bf 538}, 473 (2000)
  [arXiv:astro-ph/9911177].
\bibitem{pag07}
  L.~Page {\it et al.} [WMAP Collaboration],
  Astrophys.\ J.\ Suppl.\  {\bf 170}, 335 (2007)
  [astro-ph/0603450].
\bibitem{lew02}
  A.~Lewis and S.~Bridle,
  Phys.\ Rev.\ D {\bf 66}, 103511 (2002)
  [arXiv:astro-ph/0205436];
  A.~Lewis,
  Phys.\  Rev.\  D87, {\bf 103529} (2013) [arXiv:1304.4473].
\bibitem{guo11}
  Z.~K.~Guo, D.~J.~Schwarz and Y.~Z.~Zhang,
  JCAP {\bf 1108}, 031 (2011)
  [arXiv:1105.5916];
  Z.~K.~Guo and Y.~Z.~Zhang,
  JCAP {\bf 1111}, 032 (2011)
  [arXiv:1109.0067];
  Z.~K.~Guo and Y.~Z.~Zhang,
  Phys.\ Rev.\ D {\bf 85}, 103519 (2012)
  [arXiv:1201.1538].






\bibitem{Prunet:2004zy}
  S.~Prunet, J.~-P.~Uzan, F.~Bernardeau and T.~Brunier,
  Phys.\ Rev.\ D {\bf 71}, 083508 (2005)  [astro-ph/0406364].

\bibitem{Gordon:2005ai}
  C.~Gordon, W.~Hu, D.~Huterer and T.~M.~Crawford,
  Phys.\ Rev.\ D {\bf 72}, 103002 (2005)  [astro-ph/0509301].

\bibitem{Gordon:2006ag}
  C.~Gordon,
  Astrophys.\ J.\  {\bf 656}, 636 (2007)  [astro-ph/0607423].




\bibitem{Hirata}
C. M. Hirata,
JCAP 0909, 011 (2009) [arXiv:0907.0703[astro-ph.CO]].


\bibitem{Allahverdi:2006we}
  R.~Allahverdi, K.~Enqvist, J.~Garcia-Bellido, A.~Jokinen and A.~Mazumdar,
  JCAP {\bf 0706}, 019 (2007)  [hep-ph/0610134].

\bibitem{Qiu1} T. Qiu, J. Evslin, Y.F. Cai, M.Z. Li, X.M. Zhang, JCAP
\textbf{1110}, 036 (2011); D.A. Easson, I. Sawicki, A. Vikman,
JCAP \textbf{1111}, 021 (2011); M. Osipov and V. Rubakov,
arXiv:1303.1221; T. Qiu, X. Gao and E. N. Saridakis,
arXiv:1303.2372; D.A. Easson, I. Sawicki, A. Vikman,
arXiv:1304.3903.


\bibitem{Kanekar:2001qd}
  N.~Kanekar, V.~Sahni and Y.~Shtanov,
  Phys.\ Rev.\ D {\bf 63}, 083520 (2001)  [astro-ph/0101448].

\bibitem{Piao:2004hr}
  Y.~-S.~Piao and Y.~-Z.~Zhang,
  Nucl.\ Phys.\ B {\bf 725}, 265 (2005)  [gr-qc/0407027].

\bibitem{Ijjas:2013vea}
  A.~Ijjas, P.~J.~Steinhardt and A.~Loeb,
  arXiv:1304.2785 [astro-ph.CO].

\bibitem{V1} A. Vilenkin, Phys. Rev. \textbf{D27}, 2848 (1983); A.D. Linde, Phys. Lett. \textbf{B175}, 395 (1986); P.J. Steinhardt, in ``The Very Early Universe", ed. by G.W.
Gibbons, S.W. Hawking and S.T.C. Siklos (Cambridge University
Press, 1983).




\bibitem{GV} J. Garrige and A. Vilenkin, Phys. Rev. \textbf{D57},
2230 (1998).





\bibitem{Linde:2006nw}
  A.~D.~Linde,
 JCAP {\bf 0701}, 022 (2007)  [hep-th/0611043].



\end{thebibliography}
\end{document}